\documentclass[debug,overfull,figures,cite,bm]{epl}

\usepackage{amsmath}
\usepackage{amsfonts}
\usepackage{amsmath}
\usepackage{amssymb}
\usepackage{latexsym}
\usepackage{graphicx}

\title{Compaction dynamics of a granular medium under vertical tapping}
\shorttitle{Compaction dynamics of a granular media etc.}
\author{ P.~Philippe, D.~Bideau}
\institute{ G.M.C.M., Bat 11A, Campus de Beaulieu,
Universit\'e de Rennes I \\ 35042 Rennes, France}


\begin{document}

\maketitle

\begin{abstract}
We report new experimental results on granular
compaction under consecutive vertical taps. The evolution of the mean
volume fraction and of the mean potential energy of a granular packing
presents a slow densification until a final steady-state, and is
reminiscent to usual relaxation in glasses via a stretched exponential
law. The intensity of the taps seems to rule the characteristic time
of the relaxation according to an Arrhenius's type relation . Finally,
the analysis of the vertical volume fraction profile reveals an almost
homogeneous densification in the packing.
\end{abstract}

\section{Introduction}

Granular matter is a well known example of athermal systems; it means systems
where classical thermodynamics do not apply since thermal energy
($k_BT$) is insigni\-ficant compared to the gravitational energy of a
macroscopic grain.
A static packing of grains is therefore in a metastable state,
indefinitely trapped in a
local minimum of the total potential energy. When submitted to an
external perturbation, the system instantaneously acquires an extra mechanical
energy and then relaxes to a new metastable configuration, which
depends on the previous one as well as on the
nature of the perturbation. This dependence can be
investigated by implementing at regular intervals identical external
excitations
on an assembly of grains and analyzing the succession of static
metastable states explored by the system. This is a common experiment
in pharmaceutics when compacting powders but it is also a practical
way to ``thermalise'' a granular media and to test the conceptual
connection between granular compaction and the very slow relaxations
of out-of-equilibrium thermal systems \cite{Edwards,Nicodemi}.

The first experiments in this spirit have been carried out in Chicago
\cite{Chicago1,Chicago2,Chicago3}. Starting from a loose packing of
beads confined in a tube, a succession of vertical taps
of controlled acceleration induces a progressive and very slow compaction of
the system. This evolution is well fitted by the inverse of the
logarithm of the number of taps and, after more than 10,000 taps, a hypothetic
steady state is still not reached. These results have motivated many
theorical and numerical works, most of them dealing with the notion
of free volume and geometric constraint
\cite{Barker,Barrat,Parking}.
Some of them underscore structural aging effects, as
currently noticed in glasses.

In this paper, we present new compaction experiments in what we
believe to be more general conditions. Indeed, the previous
experiences \cite{Chicago1,Chicago2,Chicago3} were realized in a thin
cylinder of diameter $D=1.88$ cm filled with monodisperse glass
spheres of diameter $d=1$, 2, or 3 mm, that is to say a horizontal
gap of 10 to 20 beads ($N_h \sim 10$) between the lateral
walls. This condition allows a local measurement of the
volume fraction with a capacitive method and prevents any convection in
the packing. But, in return, the boundary effects are very strong and
may be in  particular responsible for the highest values of the volume fraction
obtained in some Chicago experiments\cite{Chicago2}, significantly above
the random close packing limit (approximately $64\%$)
which corresponds to the maximal volume fraction in a disordered
packing of identical hard spheres.

\begin{figure}[ht]
  \onefigure[width=8cm]{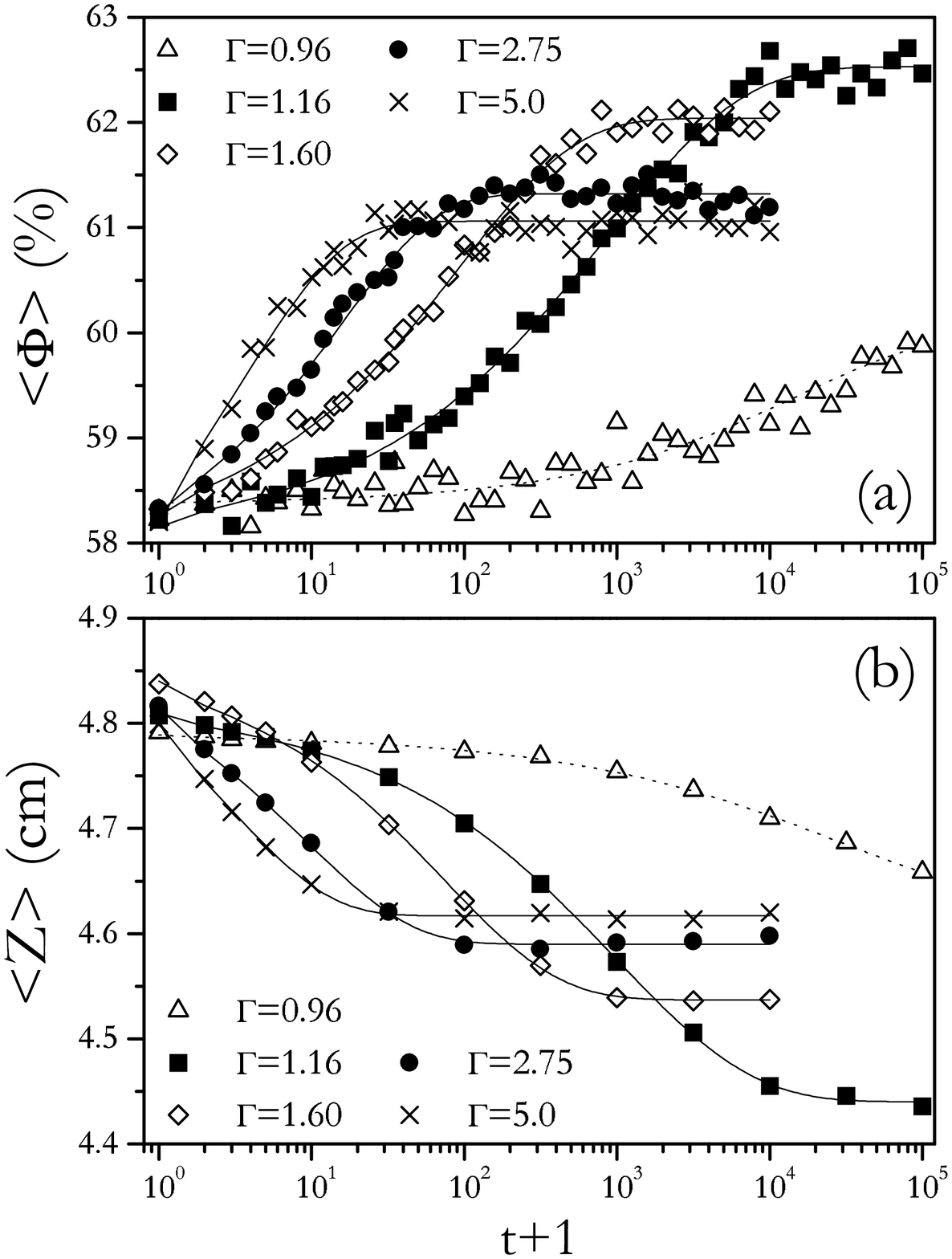}
  \caption{(a) Temporal evolution of the mean volume fraction
  $\langle \Phi \rangle$ for
  different tapping intensities $\Gamma$: $\Gamma =0.96 \ (\triangle)$,
  $\Gamma =1.16 \ (\blacksquare)$, $\Gamma =1.60 \ (\Diamond)$,
  $\Gamma =2.75 \ (\bullet)$, and $\Gamma =5.0 \ (\times)$.
  (b) Temporal evolution of the mean potential energy
  $\langle Z \rangle$ for the same values of $\Gamma$. To make the
  curves easier to examine, only an approximate $25\%$ of the
  experimental points of the mean volume fraction are plotted.}
\label{evolution}
\end{figure}

\section{Experimental set-up}

Our experimental setup is the following: a glass
cylinder of diameter 10 cm, filled with 1 mm diameter glass beads on
about 10 cm height, is shaken at regular intervals ($\Delta
t=1$ s) by an
electromagnetric exciter delivering vertical taps, each of them consisting of
an entire cycle of a sine wave (frequency $f=30$ Hz). The negative peak
acceleration $a_{max}$ felt by the whole system is measured
by an accelerometer at the bottom
of the cylinder so as to parametrize the tap intensity by the
dimensionless acceleration $\Gamma=a_{max}/g$. By measuring the
absorption of a
$\gamma$-ray beam through the packing, it is possible to estimate the
average volume
fraction in the bulk $\langle \Phi \rangle$ as well as the vertical
density profile $\Phi (z)$, providing a local analysis of the packing
structure. We can also deduce the mean potential energy
of the heap: $\langle Z \rangle = \int_0^{\infty} z\Phi (z)dz\ \big/
\int_0^{\infty} \Phi(z)dz$.\\

To restrict the boundary effects, we
use here a large horizontal gap ($N_h \sim 100$) even if,
therefore, we allow convection to occur during the compaction of the
beads, giving rise in particular to
an instability of the horizontal free surface \cite{Convection}.
In comparison with the previous experiments
\cite{Chicago1,Chicago2,Chicago3}, another important
difference concerns the vertical pressure in the static packing: in the narrow
tube used in Chicago's set-up, the pressure felt by the heap at a
given height does not correspond to the total weight of the upper
packing; part of this weight is screened by the lateral walls (``Janssen
effects''). As a consequence, in a static configuration, the vertical
pressure is homogeneous in
almost all the packing whereas, here, the vertical pressure is definitely not
saturated in the packing but probably close to the hydrostatic
pressure. Nevertheless, we do not know to what extent the initial
static situation
can play a role in the dynamical process induced by a tap.

The measure is deduced from the transmission ratio of the horizontal collimated
$\gamma$ beam through the packing: $T=A/A_0$ where $A$ and $A_0$ are
respectively the activities counted on the detector with and without
the presence of the beads in the cylinder. From the Beer-Lambert's law for
absorption, we can derive an estimation of the volume
fraction in the  probe zone: $\Phi \approx -(\mu D)^{-1} ln(T)$.
Here $\mu$ is the absorption coefficient of the beads; it was
evaluated experimentally to $\mu \approx 0.188 \ cm^{-1}$ for our
$\gamma$ beam of energy 662 $keV$ ($^{137}Cs$ source).

\begin{figure}[ht]
  \onefigure[width=8cm]{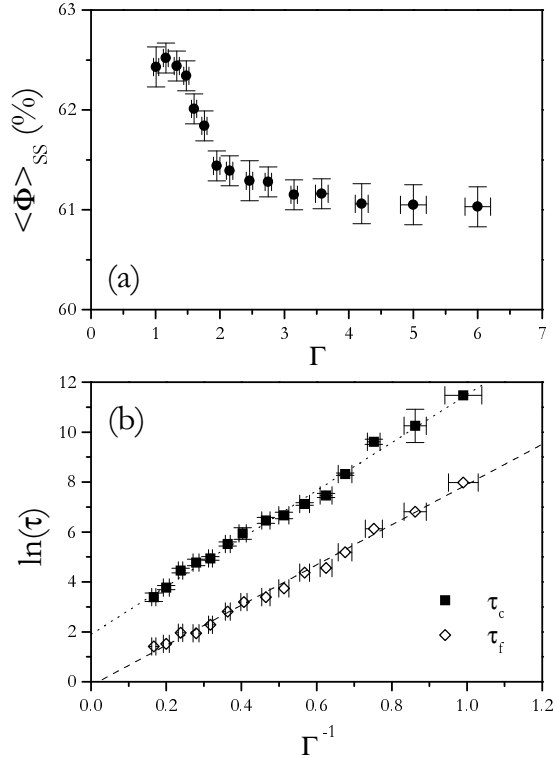}
  \caption{(a) Volume fraction of the steady-state $\langle
  \Phi\rangle_{ss}$ versus $\Gamma$.
  (b) The experimental estimation of the relaxation time
  $\tau_{c}$ ($\blacksquare$) and the characteristic time
  $\tau_{f}$ ($\Diamond$) of the
  streched exponential fit as functions of the inverse of the tapping
  intensity $\Gamma$ ($\tau_{c}$ and $\tau_{f}$ are defined in the
  text). The lines are linear fits corresponding to
  Arrhenius laws (see equation (\ref{tau})). }
\label{time}
\end{figure}

The collimated $\gamma$ beam is nearly cylindrical with a diameter
of 10 mm and
intercepts perpendicularly the vertical axis of the cylinder of beads. An
acquisition-time of 60 seconds for each measure was found as a good
consensus between
the intrinsic uncertainty of the radioactive beam and the total duration
of an experiment. We then achieve a precision $\Delta \Phi \approx 0.003$.
Two types of measurement are used: vertical profile $\Phi(z)$ and
mean volume fraction $\langle \Phi \rangle$. The vertical volume
fraction profile is deduced from 63 measures of the
$\gamma$-transmission at successive heights $z$ with a regular step
$\Delta z \approx 2$ mm; each measure is an average on a
horizontal slice obtained by 1 turn rotation of the cylinder about its
axis during
the measure. $\langle \Phi \rangle$ is estimated from the
transmission ratio $T$ averaged on
approximately 7 cm-height from the bottom of the cylinder:
the cylinder achieves a vertical
translation of 7 cm combinated to a rotation of
2 turns so as to permit a significant saving of time.
With the aim of limiting the
duration of the experiments and of avoiding redundant
information due to the very slow evolution of the system, the
measurements are spaced out in time (on a logarithmic scale)
with 2 measures of profile and 50 measures of $\langle \Phi \rangle$
per decade (except 10 for the first decade).

\section{Compaction dynamics}

Several compaction experiments were carried out for different values
of the tapping strength $\Gamma$ in the range [0,6].
Part of the results are presented in figure \ref{evolution} which shows
the evolution of the mean volume fraction $\langle
\Phi\rangle$ and of the mean potential energy  $\langle Z \rangle$
during 10,000 or 100,000 taps and for
a few values of $\Gamma$. These curves are either raw data or an
average on 2 or 3 realizations. Here we call ``time'', $t$, the number
of taps and the ``dynamics'' is the succession of static equilibrium
induced by the taps. $\langle \Phi\rangle$ and $\langle Z \rangle$
are plotted versus $log(t+1)$ as a convenient way to include the
initial state ($t=0$) on the logarithmic axis. This initial state
corresponds to a loose packing ($\langle \Phi\rangle=58.3\pm 0.3 \%$)
preparated in a reproductible way.

The typical evolution is a slow compaction of the packing
charaterised by an increase of $\langle \Phi\rangle$ and a
reciprocal decrease of $\langle Z \rangle$. Then, after a varying
relaxation time, the system finally reaches a steady-state. Note
that for $\Gamma=0.96$, the temporal window accessible in the
experiments becomes too small to observe the whole relaxation
process; but a compaction obviously occurs although $\Gamma$ is
smaller than the ``dynamical'' threshold $\Gamma^*$
($\Gamma^*\approx 1.2$) above which there is a collective takeoff
of the packing from the bottom of the cylinder. When reached, the
final steady-state is all the more compact (i.e. small value of
$\langle Z \rangle_{ss}$ and high value of $\langle
\Phi\rangle_{ss}$) as the tapping strength is slight (see figure
\ref{time}a); but, in return, the number of taps needed increases
significantly. In all our experiences, the volume fraction stays
below the random close packing limit which tends to reject any
hypothesis of ordering or crystallization in the packing.
Convection seems also to play a role in the compaction process: on
figure \ref{time}a, we observe indeed a significant change in the
dependance of $\langle \Phi\rangle_{ss}$ with $\Gamma$ which might
correspond to different convective regimes. Under a threshold
$\Gamma_c \approx 2$, the final state of the free surface of the
packing is an inclined plane and indicates a spontaneous breaking
of symmetry. Above $\Gamma_c$, the free surface heaps up
moderately and finally takes a flat conical shape probably brought
about by a nearly torical convective roll. These sorts of free
surface instabilities have already been observed, see for instance
\cite{Convection}.

\begin{figure}[ht]
  \onefigure[width=9.5cm]{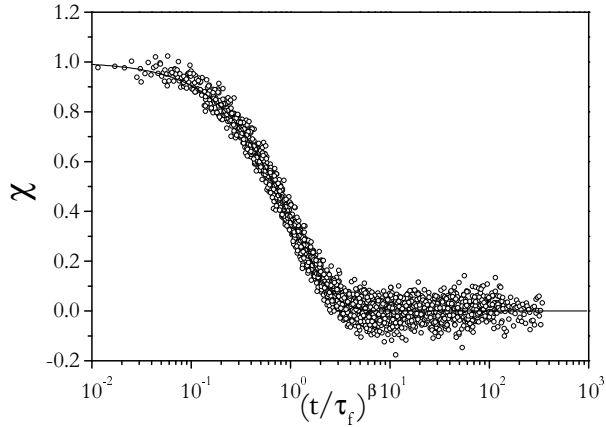}
  \caption{Collapse of the compaction curves obtained with
  $\Gamma=1.01$, $1.16$, $1.33$, $1.48$, $1.60$, $1.76$, $1.95$, $2.15$,
  $2.46$, $2.75$, $3.15$, $3.58$, $4.20$, $5.0$, and $6.0$:
  $\chi=(\langle
  \Phi\rangle_{ss}-\langle \Phi\rangle(t))/(\langle
  \Phi\rangle_{ss}-\langle \Phi\rangle(0))$ is plotted versus
  $(t/\tau_f)^{\beta}$. The solid line is the function $f(u)=exp(-u)$
  corresponding to the stretched exponential law.}
\label{collapse}
\end{figure}

In comparison with the previous experimental results
\cite{Chicago1}, some sharp differences appear. First, the shapes of
the compaction curves differ significantly, especially concerning the
long time behavior and the
obtaining of a final steady-state which is here definitely
established and may correspond to a dynamical balance between
compaction and convection. Moreover, for a given
intensity $\Gamma$, the dynamics of the compaction seems slower in the
Chicago's experiments, particularly for the highest values of
$\Gamma$. We believe that these differences are principally due to the
disparity of the lateral constraint (horizontal gap $N_h$) in the
two configurations ($N_h \sim 10$ against $N_h \sim 100$).
To extend the comparison, we have tried to fit $\langle \Phi\rangle$ with the
empirical law initially proposed in \cite{Chicago1}:
\begin{equation}
\langle \Phi\rangle(t) =\langle \Phi\rangle_{\infty} -
\frac{\Delta \langle \Phi\rangle_{\infty}}{1+B\ln(1+{t/\tau})}
\label{fit-chicago}
\end{equation}

The result is satisfactory concerning the beginning of a typical
compaction curve but
fails to correctly fit the final relaxation up to the steady-state.
In particular, the parameter $\langle \Phi\rangle_{\infty}$
overestimates sharply the steady-state volume fraction $\langle
\Phi\rangle_{ss}$ at small $\Gamma$. On
contrary, our datas concerning $\langle \Phi\rangle$ as well as $\langle
Z\rangle$ are in very good agreement with a stretched
exponential function (equation(\ref{KWW})) on the whole temporal
range. First used by
Kohlrausch in 1854 \cite{Kohlrausch}, this expression was far later
popularized by Williams and Watts \cite{WW}. It is now frequently applied
to a large range of relaxations in
disordered thermal systems as glasses (see for example \cite{Phillips} and
references therein) and is often called KWW law.
\begin{equation}
X(t) =X_{\infty} - (X_{\infty}-X_0)
\exp\big(-(t/\tau_f)^{\beta}\big) \ \ \ \ \text{with $X=\langle \Phi\rangle$ or $\langle Z\rangle$}
\label{KWW}
\end{equation}

As $X_{\infty}$ can be approximated by  $X_{ss}$ and $X_0$ by
$X(t=0)$, equation (\ref{KWW}) has only two free parameters ($\tau_f$
and $\beta$). Examples of
this fit are presented in figure \ref{evolution} where the
solid lines are the stretched exponential laws. For
$\Gamma=0.96$ and more generally for $\Gamma \lesssim 1$, only the
beginning of the relaxation is accessible and there is so no evidence
that a stretched exponential law still describes the compaction
dynamics; consequently the corresponding fit is plotted in dotted
line. The values obtained for
$\beta$ are in the range $0.5-0.8$ and tend to increase slightly with $\Gamma$
whereas $\tau_f$ decreases strongly with $\Gamma$.

A general collapse of
the compaction curves can be obtained by use of the function $\chi$,
defined as the rate of increase of $\langle \Phi\rangle$:
$\chi(t)=\big(\langle \Phi\rangle_{ss}-\langle
\Phi\rangle(t)\big)\big/\big(\langle \Phi\rangle_{ss}-\langle
\Phi\rangle(0)\big)=f\big((t/\tau_f)^{\beta}\big)$.
Figure \ref{collapse} presents the plot of $\chi$ versus
$(t/\tau_f)^{\beta}$; the solid line plots the KWW fit. The
fluctuations at large $t$ correspond to the final steady-state and are
particularly significant at large $\Gamma$.

\section{Relaxation time}

To quantify more precisely the influence of $\Gamma$
on the dynamics of the compaction, it is possible to estimate a
characteristic relaxation time. To do this, we use two different initial
packings: the loose one, already presented above
($\langle \Phi\rangle=58.3\pm 0.3 \%$), and a
more compact one ($\langle \Phi\rangle=63.2\pm 0.2 \%$). Submitted
to identical taps, the first packing densifies whereas the other dilates and
both of them progressively meet, reaching the same steady-state. From
this meeting point, we can evaluate a time of convergence $\tau_{c}$.
As most of the usual disordered packings of
monodisperse beads have a packing fraction comprised between $58.3\%$
and $63.2\%$, this time can also be
regarded as a memory effect and can be interpreted as the number of
taps required for a packing
to ``forget'' its initial configuration. When analysing the dependance
with $\Gamma$ of this characteristic time $\tau_{c}$ as well as of the
parameter time
$\tau_{f}$ of the stretched exponential fit, we found that an
Arrhenius behavior (equation(\ref{tau})) describes reasonably
well the experimental dynamics as illustrated in figure \ref{time}b:
\begin{equation}
\tau_{c,f}(\Gamma)=\tau_0 \exp(\frac{\Gamma_0}{\Gamma})
\label{tau}
\end{equation}
In the two cases, we have obtained
($\Gamma_0 \approx 9.6$, $\tau_0 \approx
6.7$) for $\tau_{c}$ and ($\Gamma_0 \approx 8.0$, $\tau_0 \approx 0.9$)
for $\tau_{f}$.

\begin{figure}[ht]
  \onefigure[width=10cm]{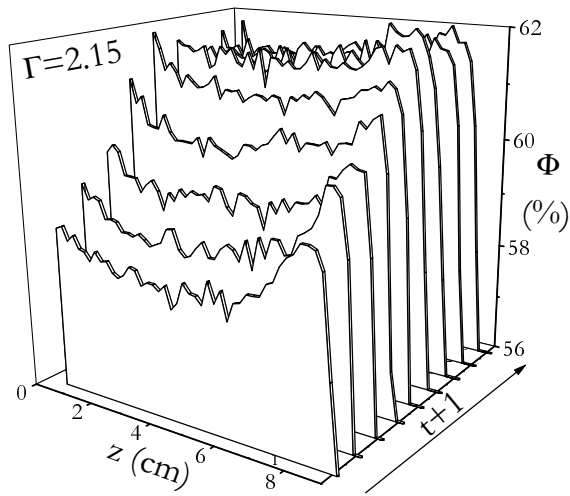}
  \caption{Quasi-homogeneous compaction of the vertical volume fraction
  profile. The profiles are represented in the range $56\% < \Phi
  <62\%$ and $0 < z < 9$ cm and correspond to the following tap
  numbers: $t+1=1$, $3$, $10$, $32$, $100$, $316$, $1.000$, $3.162$, and
  $10.000$.}
\label{profils}
\end{figure}

Finally, we present in figure \ref{profils} a typical evolution of
the vertical volume fraction profile in the case of $\Gamma=2.15$.
To observe the progressive
densification of the packing, we use a zoom on the zone of interest
($56\% < \Phi < 62\%$) and a three-dimensional representation with the
third axis corresponding to time on a logarithmic scale. The
profile is plotted approximately every half-decade.

We note that the
compaction is rather homogeneous in the bulk; there is no upward or
downward densification front. The profile continuously reaches a final
steady-state. This asymptotic profile is nearly uniform, a
slight positive gradient ($d\Phi / dz >0$) appears for $\Gamma \geq 2$;
this observation agrees with previous experimental \cite{Chicago1}
and numerical \cite{Barrat}
results. We have also verified that the same steady-state profile is
obtained when starting from the initial dense packing ($\langle
\Phi\rangle=63.2\pm 0.2 \%$) instead of the loose one.

\section{Conclusion and perspectives}

In conclusion, we have shown that, in the case where the lateral
constraint is weak, granular compaction by vertical taps
is quite similar to a typical relaxation of a out-of-equilibrium
thermal system. Indeed, the densification curves can be
reasonably well fitted by a stretched exponential or KWW law. The
characteristic time of compaction ($\tau_c$ or $\tau_f$) follows an
Arrhenius relation where the dimensionless acceleration $\Gamma$ plays the
role of temperature. These
results confirm and reinforce the analogy between compaction dynamics
and ``glassy'' phenomena. Moreover,
the acquisition of the vertical volume fraction profile has
established that the compaction is rather homogeneous in all the
packing. Further investigations are in progress to study more
precisely the influence of the lateral constraint ($N_h$) and to
clarify the link between convection and compaction.\\

\acknowledgements

We are grateful to A. Valance and R. Delannay for a careful reading of
the manuscript, and to S. Bourles and P. Chasle for technical assistance.

\end{document}